\documentclass[osajnl,preprint,showpacs,superscriptaddress,12pt]{revtex4-1} %% use 12pt for preprint option
\usepackage{amsmath,amssymb,graphicx}
\begin{document}

\title{Quantitative interpretation of time-resolved coherent anti-Stokes Raman spectroscopy with all Gaussian pulses}

\author{Gombojav O. Ariunbold}
\affiliation{Department of Physics and Astronomy, Mississippi State University, Starkville, MS 39762, USA}

\author{Narangerel Altangerel}
\affiliation{Department of Physics and Astronomy, Texas A${\rm \&}$M University, College Station, Texas 77843, USA}

\begin{abstract}
Coherent Raman scattering spectroscopy is studied purposely, with the Gaussian ultrashort pulses as a hands-on elucidatory extraction tool of the clean coherent Raman resonant spectra from the overall measured data contaminated with the non-resonant four wave mixing background. The integral formulae for both the coherent anti-Stokes and Stokes Raman scattering are given in the semiclassical picture, and the closed-form solutions in terms of a complex error function are obtained. An analytic form of maximum enhancement of pure coherent Raman spectra at  threshold time delay depending on bandwidth of probe pulse is also obtained. The observed experimental data for pyridine in liquid-phase are quantitatively elucidated and the inferred time-resolved coherent Raman resonant results are reconstructed with a new insight.
\end{abstract}

%\ocis{(140.3490) Lasers, distributed-feedback; (060.2420) Fibers, polarization-maintaining; (060.3735) Fiber
%Bragg gratings; (060.2370) Fiber optics sensors.}% REPLACE WITH CORRECT OCIS CODES FOR YOUR ARTICLE
                          % NOTE: \ocis{} IS ALIASED TO \pacs{} BUT MUST
                          % FORMAT THE TERMS CORRECTLY FOR EACH JOURNAL

\maketitle %% required

%\section{Introduction}
%
Coherent anti-Stokes Raman scattering is a powerful spectroscopic technique. Since its first demonstration~\cite{Terhune}, recent significant developments of the detection of bacterial spores~\cite{AriScience,AriPNAS}, coherent Raman microscopy for biological tissues~\cite{Duncan,Zumbush}, gas-phase thermometry of reacting and non-reacting flows~\cite{Anil} and many others (see, recent reviews~\cite{Xie book,Camp,Austin,Kraft1, Mogilevsky, Silberberg, Zheltikov,Tolles}) have been the state-of-art success of this fascinating technique. Implementation of ultrafast laser pulses is particularly the key ingredient here~\cite{Scully}. In the ultrafast coherent Raman scattering process, two pulses (pump and Stokes) resonantly excite molecular Raman vibrations in unison and a third pulse (probe) then scatters off from these coherently vibrating molecules. The scattered light in coherent anti-Stokes Raman scattering (CARS) is blue-shifted as opposed to the red-shifted light in coherent Stokes Raman scattering (CSRS). 
However, the overall measured spectral data are contaminated with the non-resonant four wave mixing background. Among various techniques to overcome this four wave mixing (FWM), CARS experiments with broadband (fs) pump and Stokes excitation pulses with a narrowband (ps) shaped and delayed probe pulse have been successfully demonstrated~\cite{AriScience,AriPNAS, AriOL, Hamaguchi, Prince, Stauffer1, Stauffer2, Selm, Konorov, Marangoni, Jiahua}. Time-resolved CARS spectrogram data are a collection of spectra recorded at each probe pulse time delay (for instance, see~\cite{AriScience,AriPNAS,AriOL}). 
By delaying probe pulse with respect to the excitation pulses, one can reveal the rich time-resolved feature of the CARS process, in addition to background-suppressed CARS spectra selected at fixed optimal time delay within the spectrogram data. 
The efficient FWM eliminations have been recently demonstrated with probe pulses shaped in time- and frequency-domain as in sinc-square~\cite{AriScience,AriOL,AriPNAS,Prince,Stauffer1,Stauffer2}, square-sinc~\cite{Selm}, exponential-Lorentzian~\cite{Marangoni, Stauffer1, Stauffer2, Konorov} and Gaussian-Gaussian~\cite{Jiahua} forms.
Particularly, experiments on the impulsive CARS, where a common broadband excitation pulse that contains both pump and Stokes portions with degenerate center frequencies accompanied with a delayed narrowband probe pulse, have been reported in~\cite{Selm, Marc Lee, Jiahua}, and
its approximate closed-form solutions with all above-mentioned pulse shapes have been obtained in~\cite{Marrocco1,Marrocco2}.  
However, an accurate extraction of the clean coherent Raman spectra without FWM contamination for various pulse shapes in the entire time delay range still heavily depends on trusting numerical efforts, usually without proper interpretations of the actual ongoing processes. Therefore, a simple deductive tool with a sufficient quantitative interpretation for the time-resolved coherent Raman scattering experiments is on demand.

In this letter, we study the time-resolved CARS and CSRS with a Gaussian-Gaussian probe pulse and  obtain the exact closed-form solutions beyond impulsive excitation. The obtained simple solutions with Gaussian pulses can interpret the main features of coherent Raman spectroscopy without even performing rigorous simulations with other non-Gaussian sophisticated probe pulses. The present letter is organized as follows. First, a suitable theoretical description of CARS/CSRS signals is given.
Then, the exact solutions for Gaussian pulses are derived. Finally, the time-resolved CSRS experimental data are explained, followed by a conclusion.
%

%\section{General approach}
%
Let us start with a theoretical description of the third-order nonlinear process.
The overall time-resolved signal
is expressed as a sum of the FWM and CARS or CSRS signals as $| P^{(3)}(\omega_{aS,S},\tau)|^2=| P_{FWM}^{(3)}(\omega_{aS,S},\tau)+P_{CARS,CSRS}^{(3)}(\omega_{aS,S},\tau)|^2$ where the explicit integral form for the FWM signal is given by $P_{FWM}^{(3)}(\omega_{aS,S},\tau)  =
c_0 \int_{-\infty}^\infty d\omega  { E}_{3}(\delta_{aS,S}\mp\omega,\tau) S_{aS,S}(\omega)$ and for the CARS/CSRS it is obtained as
\begin{eqnarray} \label{AriFormula}
P^{(3)}_{CARS,CSRS}(\omega_{aS,S},\tau)  = 
%\nonumber\\
%\\
 c_{1,2} \int_{-\infty}^\infty  
d\omega \frac{{ E}_3(\delta_{aS,S}\mp\omega,\tau) }{\Delta-\omega \mp i\Gamma/2}
S_{aS,S}(\omega) %\nonumber
\end{eqnarray}
Here $E_{1,2,3}$ are electric fields for pump, Stokes and probe ultrashort pulses with the corresponding center frequencies $\omega^0_{1,2,3}$, whereas $\omega_{aS,S}$ and $\tau$ are the frequencies of the anti-Stokes and Stokes pulses and delay of the probe pulse.
$S_{aS}(\omega)  = \int_{-\infty}^\infty d\omega'{{ E}_1(\omega'){E}_2^*(\omega'-\omega)}$
and $S_{S}(\omega)  = \int_{-\infty}^\infty d\omega'{{ E}_1(\omega'+\omega){E}_2^*(\omega')}$ are the convolutions.  $c_{1,2}$ and $c_0$ are constant complex coefficients, and also an extra common term ${\rm exp}( i\omega_3^0\tau)$ is omitted here. The other parameters are $\Delta=\Omega_R - (\omega_1^0-\omega_2^0)$ (Raman detuning with $\Omega_R$ being Raman vibrational frequency), $\omega_{aS,S}^0 = \omega_3^0 \pm \Delta\omega^0$ (center frequencies of anti-Stokes or Stokes signal) and $\delta_{aS,S} = \omega_{aS,S}-\omega_{aS,S}^0$ (detunings for CARS or CSRS signal). A rigorous CARS/CSRS theory has been developed in~\cite{He book, Mukamel book, Xie book, Tehver, Yuratich, Zheltikov, Carreira, Oron, Dudovich}. In derivation of Eq.(\ref{AriFormula}), we assume that phase matching condition is satisfied and any propagation effect is not included here for the sake of simplicity without losing the essential ingredients of the coherent Raman processes. The above formula Eq.(\ref{AriFormula}) is, indeed, in its most convenient form because the integration variables and their limits are carefully chosen here. The CARS formula has been derived in~\cite{Marrocco1,Marrocco2} as time-delayed Yuratich equations~\cite{Yuratich} for the first time to our knowledge, like the formula for the CSRS in the present letter. Multiple Raman resonances can be included simply as a sum of the multiple Lorentzians in Eq.(\ref{AriFormula}) with the corresponding coefficients $c_{1,2}$.
%

%\section{all Gaussian exact solutions}
%
In the following, we present the exact solutions for Gaussian pump, Stokes and probe pulses.
With this assumption, the complicated integral formula is straightforwardly evaluated to achieve closed-form solutions. The CARS/CSRS solutions of Eq.(\ref{AriFormula}) are found to be
\begin{eqnarray}   \label{AriSolutions}
&&{P^{(3)}_{CARS,CSRS}(\omega_{aS,S},\tau)}=
\nonumber\\
&&\mp i c_{1,2}'
W_{12} w(\zeta_{\pm})
{\rm e}^{-\frac{\tau^2}{2t_{FWM}^2}}
{\rm e}^{- 2ln2\frac{\delta_{aS,S}^2}{\Delta\omega_{123}^2}}
 {\rm e}^{i\delta_{aS,S}\frac{\Delta\omega_3^2}{\Delta\omega_{123}^2}\tau}
\end{eqnarray}
where the FWM solution is obtained as $P_{FWM}^{(3)}(\omega_{aS,S},\tau) = c_0' W_{12} W_{123}$ $
{\rm exp} (-{\tau^2}/{2t_{FWM}^2}) $ ${\rm exp}(- 2ln2{\delta_{aS,S}^2}/{\Delta\omega_{123}^2})$ $ {\rm exp}(i\delta_{aS,S}\tau{\Delta\omega_3^2}/\Delta\omega_{123}^2)$.
The parameters here include $c_{0}'={\pi c_{0} A_1A_2^*A_3}/{4ln2}$ and $c_{1,2}'={\pi^{3/2} c _{1,2}A_1A_2^*A_3}/\sqrt{8ln2}$ (modified constant coefficients with the input pulses amplitudes $A_{1,2,3}$), $\Delta\omega_{1,2,3}$ (spectral full width at half maxima (FWHM) of the input pulses) and the rest involves $W_{12}^2 = \Delta\omega_1^2 \Delta\omega_2^2 / (\Delta\omega_{1}^2 + \Delta\omega_2^2)$, $W_{123}^2 = (\Delta\omega_{1}^2 + \Delta\omega_2^2)  \Delta\omega_3^2/ (\Delta\omega_{1}^2 + \Delta\omega_2^2 + \Delta\omega_3^2)$. 
The FWM solution is constructed from a product of Gaussian functions with the widest spectral width, $\Delta\omega_{FWM}=(\Delta\omega_{1}^2 + \Delta\omega_2^2 + \Delta\omega_3^2)^{1/2}$, and longest time duration, $t_{FWM}=\sqrt{4ln2}/W_{123}$,
among the input three pulses. In this case, the width of the pump/Stokes pulse is the widest, and probe pulse duration is the longest. This means that the FWM background can be efficiently removed when the probe pulse is delayed at a time longer than (or comparable to) its duration. 
The CARS/CSRS solutions are written in terms of the Faddeeva function~\cite{Faddeeva,Abrarov,Ali} $w(\zeta_{\pm})$ or the error function with a complex argument given as 
$\zeta_\pm = [(\delta_{aS,S} W_{123}^2/\Delta\omega_3^2 \mp \Delta + i \Gamma/2)t_{FWM} - i \tau/t_{FWM}]/\sqrt{2}$ 
where $\Gamma$ is FWHM of a chosen Raman line.  A special solution of the impulsively excited CARS in the limit $W_{123} \rightarrow \Delta\omega_3$ has been partially derived in~\cite{Marrocco1,Marrocco2}.  This limited condition can be satisfied for the extreme broadband pump/Stokes and narrowband probe pulses. The extended  solution for the multiplex CARS covering multiple Raman resonances is simply a superposition of the Faddeeva functions as $\sum_{j=1} c_{1,2,j}' w(\zeta_{\pm,j})$, where the $j$th term is defined as
$\zeta_{\pm,j} = [(\delta_{aS,S} W_{123}^2/\Delta\omega_3^2 \mp \Delta_j + i \Gamma_j/2)t_{FWM} - i \tau/t_{FWM}]/\sqrt{2}$. 
The FWM, CARS and sum spectra for a single Raman line are compared at zero and positive delay of the probe pulse. 
At zero delay, in the sum spectra the FWM distorts pure CARS signal in the experiment. The extraction method based on Kramers-Kronig relation has been developed~\cite{Marc Liu} and implemented, for instance, in~\cite{Marc Camp}. It is trivial that as probe pulse is delayed, the CARS signal decreases, like FWM. 
However, it is unexpected and even seemingly counterintuitive that the CARS signal is enhanced, as probe pulse is delayed at a certain threshold time. An enhancement of the CARS/CSRS is one of the main results of the present letter. 
\begin{figure}[htbp]
\centering
\fbox{\includegraphics[width=\linewidth]{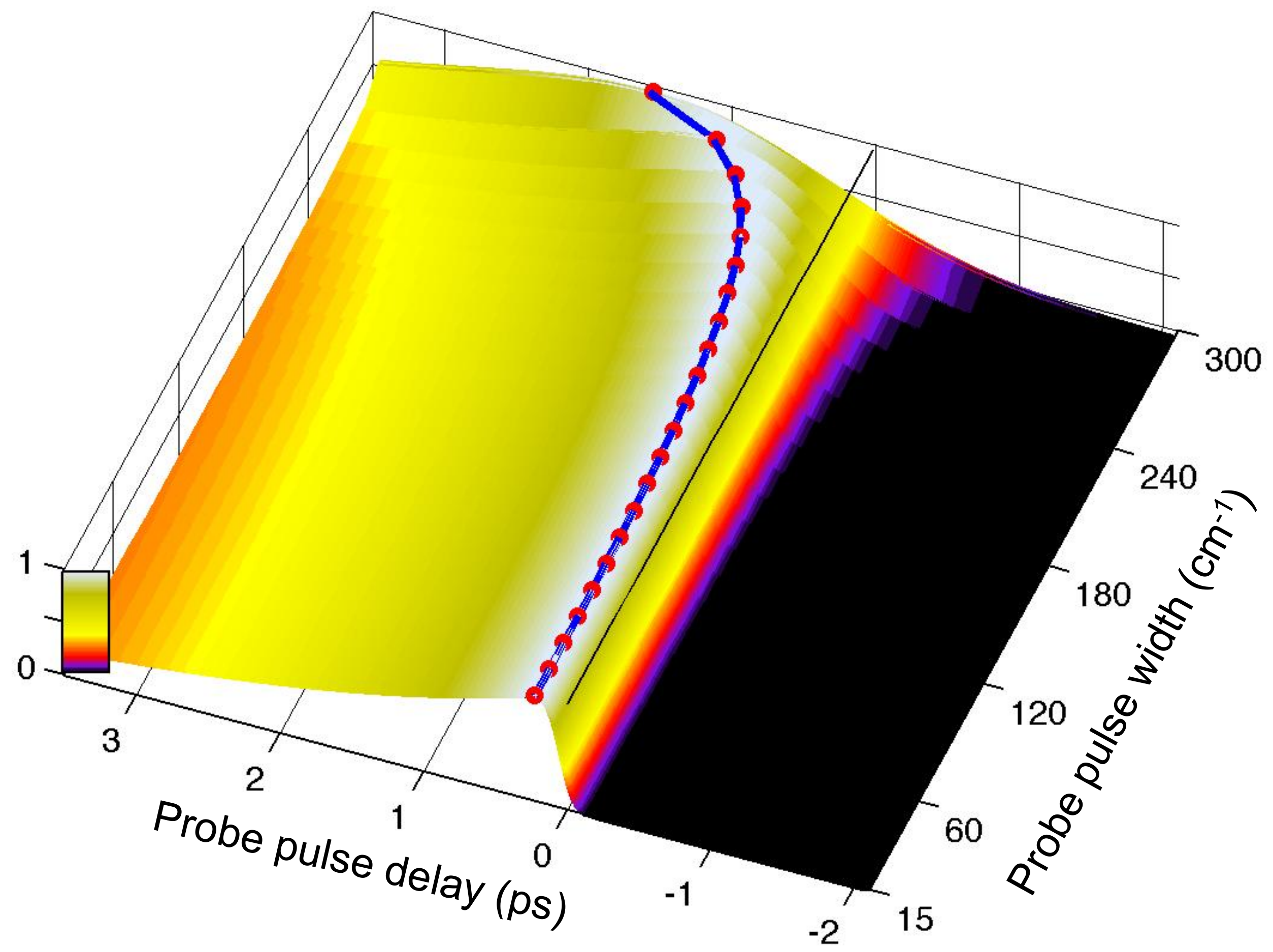}}
\caption{(Color online) The three dimensional normalized CARS theoretical result with varying probe pulse width and delay. Its maximum enhancement as a function of the threshold delay (solid curve) is accurately pinned by an analytic formula Eq.(\ref{AriTau}) (dots), escaping from zero delay (straight line).}
\end{figure}
In Fig.~1, this enhancement at a certain positive threshold time delay is studied as a function of probe pulse width. The effect is more obvious when the probe pulse width becomes narrower. A formula for the threshold time delay is found to be 
\begin{equation} \label{AriTau}
\tau_{threshold} (\Delta\omega_3)= t_{FWM}\left[ \frac{1}{\varepsilon(\Delta\omega_3,\Gamma)}+ \frac{\Gamma t_{FWM}}{2}\right]
\end{equation}
where coefficient $\varepsilon$ linearly depends inversely on the probe pulse width $\Delta\omega_3$ as $\varepsilon=a \Gamma /\Delta\omega_3 + b$ with $a=1.8$ and $b=0.38$. Threshold time delay for maximum enhancement (solid curve) and values obtained from Eq.(\ref{AriTau}) (dots) are shown in Fig.~1. It is worth noting that although enhancement at positive probe delay has been briefly mentioned previously in our work~\cite{AriPNAS} and also in~\cite{Marrocco1,Roy2,Kliewer}, an explicit accurate formula Eq.(\ref{AriTau}) is obtained here for the first time.

%\section{Time-resolved CSRS with Double Raman lines}

Next, we compare the CSRS results to the experimental data obtained for pyridine in liquid-phase. From the view point of the obtained solutions Eq.(\ref{AriSolutions}), an important deviation between the CSRS and CARS signals is either the subtraction or addition of the Raman detuning inside the argument of the Faddeeva function as $\zeta_\pm = [(\delta_{aS,S} W_{123}^2/\Delta\omega_3^2 \mp \Delta + i \Gamma/2)t_{FWM} - i \tau/t_{FWM}]/\sqrt{2}$ . Moreover, a sign difference in Eq.({\ref{AriSolutions}}) suggests that the CSRS and CARS signal amplitudes are out of phase. 
Comparison between CSRS and CARS has been studied, for instance, in~\cite{Tehver, Carreira}. A slight deviation between simultaneously recorded CARS and CSRS signals has been also reported in~\cite{Jiahua}, but otherwise, time-resolved CSRS and CARS are quite the same. 
Fig.~2 illustrates one-to-one comparison between the experimental and theoretical data.
The experimental data, part of which has been previously reported in our work~\cite{AriOL} were taken for pyridine in liquid-phase by using the Coherent Inc., Ti-sapphire amplified kilohertz laser system including the optical parametric amplifiers. The excitation pump and Stokes pulses were centered in the middle of a pair of pyridine molecular vibrational modes.These Raman modes are in-plane ring-bend $1030$ cm$^{-1}$ and ring-breathing $991$ cm$^{-1}$ separated by $39$ cm$^{-1}$ the corresponding beating period of $0.855$ ps and both the lines have similar spectral width ($\approx 2.2$) cm$^{-1}$  and Raman cross-sections~\cite{Kaiser1}.  Their dephasing lifetimes are about 2.55 ps with a slight deviation ($\approx 20\%$)~\cite{Kaiser2}. Three different probe pulse widths: $300$ cm$^{-1}$ (in the top left), $60$ cm$^{-1}$ (in the middle left) and $15$ cm$^{-1}$ (in the bottom left) were selected. For the broadband probe, the two Raman lines are unresolved, thus exhibiting a beating in time-domain, whereas, for narrowband probe, these lines are well resolved. 
\begin{figure}[htbp]
\centering
\fbox{\includegraphics[width=\linewidth]{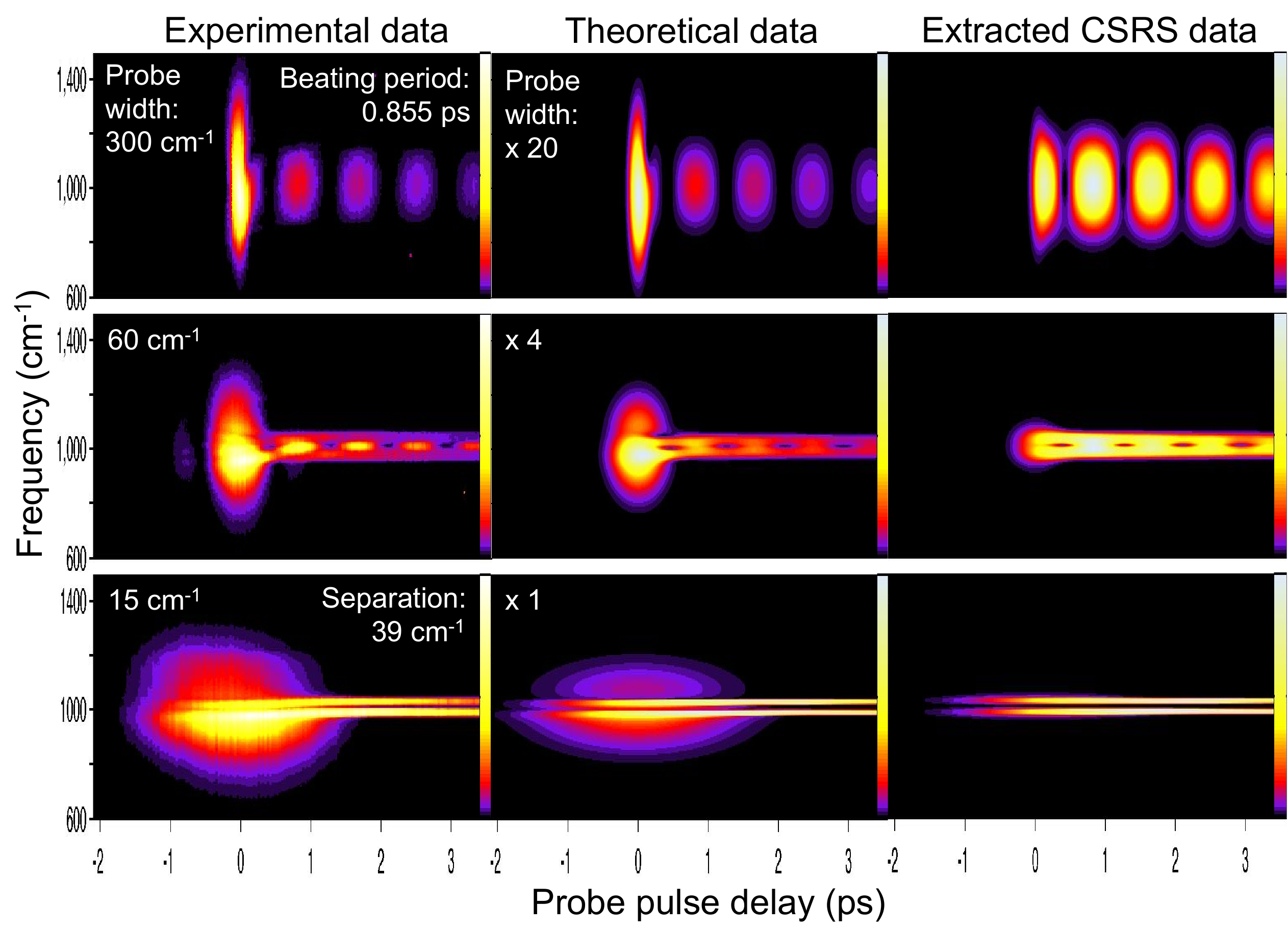}}
\caption{(Color online) The experimental data (left column), theoretical sum spectral data (center column), and extracted CSRS data (right column). All data are normalized. The experimental data were taken with probe pulse widths $300$  cm$^{-1}$, $60$ cm$^{-1}$ and $15$ cm$^{-1}$. A separation of Raman lines of pyridine molecules is $39$ cm$^{-1}$  with a beating period of $0.855$ ps.}
\end{figure}
The normalized theoretical results obtained by Eq.(\ref{AriSolutions}) with and without FWM  are shown in the next two columns. As mentioned previously, the width and duration of FWM change accordingly with probe pulse time duration as well as excitation spectral width (not shown here). 
In the theory, the narrowband probe pulse width is increased by 1, 4, and 20 times to be consistent with our experimental data. It is chosen, in addition, that the two Raman line-widths differ by 20 $\%$ and a ratio between the Raman line-width and separation is equal to the ratio of 2.2 and 39 cm$^{-1}$.
The theoretically inferred, clean CSRS results are depicted in the right column of Fig.~2, also for three different probe widths. Maxima of the FWM obviously remain at zero delay, though this is not the case for the CSRS for the narrowband probe. At the threshold time delay which is longer than the time duration of FWM, the FWM is drastically suppressed and, at the same time, the CSRS signal is enhanced. In the experiment, probe pulse shape is not strictly Gaussian, rather it is a sinc-square pulse in time- and frequency-domain. However, the theoretical results for the Gaussian probe pulse nicely fit the experimental data. This is expected for two main reasons.  First, in the experiment, a narrowband probe pulse was not ideally sinc-square shaped, rather it was a quasi-Gaussian pulse. Second, the central lobe of sinc function is the dominant part, thus it can be approximated with a Gaussian form. It is worth mentioning that a fine-tuning effect such as additional background subtraction at the optimal probe delay exactly on the first leading node of sinc function~\cite{AriScience,AriOL,Prince,Hamaguchi} is, of course, not seen in the theory. 
However, the detailed understanding of the existence of CARS/CSRS enhancement, is indeed, the most crucial part, especially for molecules such as pyridine with long lasting molecular vibrations.

In conclusion, we obtained exact solutions in terms of the Faddeeva function for CARS/CSRS spectroscopy.
We showed that the time-resolved coherent Raman spectroscopic experiments can be quantitatively elucidated with all Gaussian input laser pulses.  This is a simple interpretative tool to unearth buried pure coherent Raman spectra within the non-resonant background contamination. The efficient enhancement of coherent Raman scattering at certain threshold time delay of probe pulse with a variable spectral width has been obtained. This phenomenon adds on to the previous understanding of robust and superior background-suppressed coherent anti-Stokes Raman spectroscopy.

We would like to thank M. Cicerone, C. Camp Jr, Y.J. Lee, M. Zhi, D. Pestov, V. Yakovlev and A. Sokolov for their fruitful discussions.

%Manual citation list

\end{document}